\documentclass[traditabstract]{aa}
\usepackage{graphicx,color} 
\usepackage{rotating,subfigure,amssymb,afterpage} 
\usepackage{txfonts}

\begin{document} 
\title{Statistics and implications of substructure 
detected in a representative sample of X-ray clusters}
\author{Gayoung Chon\inst{1}, Hans B\"ohringer\inst{1}, 
Graham P. Smith\inst{2}}
\offprints{Gayoung Chon, gchon@mpe.mpg.de} 
\institute{$^1$ Max-Planck-Institut f\"ur extraterrestrische Physik, 
85748 Garching, Germany\\
$^2$ School of Physics and Astronomy, University of Birmingham, 
Edgbaston, Birmingham B15 2TT, England\\
} 
\date{Submitted 21 August 2012;accepted 16 October 2012} 

\abstract{
We present a morphological study of 35 X-ray luminous galaxy clusters
at $0.15<z<0.3$, selected in a similar manner to the Local Cluster
Substructure Survey (LoCuSS), for which deep \emph{XMM-Newton} observations
are available. We characterise the structure of the X-ray surface
brightness distribution of each cluster by measuring both their
power ratios and centroid shift, and thus rank the clusters by the 
degree of substructure. These complementary probes give a
consistent description of the cluster morphologies with some well
understood exceptions. We find a remarkably tight correlation of 
regular morphology with the occurrence of cool cores in clusters. 
We also compare our measurements of X-ray
morphology with measurements of the luminosity gap statistics and
ellipticity of the brightest cluster galaxy (BCG), to examine recent
suggestions that these quantities may be efficient probes of the
assembly history of and observer's viewing angle through
cluster-scale mass distributions. 
Our X-ray analysis confirms that cluster with large luminosity gaps 
form a relatively homogeneous population that appears more regular 
with the implication that such systems did not suffer from recent 
merger activity. 
Similarly, we find a clear correlation between the ellipticity 
of the BCGs and the shape of the cluster. In particular nearly 
circular BCGs ($\epsilon\textless0.2$) are found in undisturbed systems 
with regular X-ray morphologies. For these systems it has been suggested 
that they are intrinsically prolate and viewed along the line 
of sight close to the major axis. Finally, we check how our new
X-ray morphological analysis maps onto cluster scaling relations,
finding that (i) clusters with relatively undisturbed X-ray
morphologies are on average more luminous at fixed X-ray temperature
than those with disturbed morphologies, and (ii) disturbed clusters
have larger X-ray masses than regular clusters for a given temperature
in the $M_X-T$ relation. We also show that the scatter in the ratio of
X-ray and weak lensing based cluster mass measurements is
larger for disturbed clusters than for those of more regular
morphology. Overall, our results demonstrate the feasibility of
assembling a self-consistent picture of the physical structure of
clusters from X-ray and optical data, and the potential to apply
this in the measurement of cosmological cluster scaling relations.
} 
\authorrunning{Chon et al.}  
\titlerunning{Substructure studies of X-ray Clusters}
\maketitle 

\section{Introduction}

Galaxy clusters are ideal probes for tracing the large-scale cosmic
matter distribution and for testing cosmological models
(e.g. Vikhlinin et al. 2009, B\"ohringer 2011, Allen et al. 2011). A
prerequisite for these studies is a good measure of the galaxy cluster
masses. Only for small samples will the masses be determined
individually while for the large cosmological samples of galaxy
clusters the masses will be estimated through the application of mass
- observable scaling relations, where the observable is for example
the X-ray luminosity in the case of galaxy cluster samples from X-ray
surveys (e.g. Pratt et al. 2009, Vikhlinin et al. 2009, for more
references see B\"ohringer et al. 2012). This then requires the
calibration of the mass - observable relation with mass determinations
for a calibration sample.

X-ray observations can be used effectively to characterise the
structure of galaxy clusters, which are to first order described in
theory as dark matter halos (e.g. Navarro, Frenk and White 1995). The
X-ray emitting gas is filling the entire dark matter halos providing
an outline of their structure and a measure of their gravitational
potentials. Assuming that the X-ray emitting intracluster medium (ICM)
is filling the potential in a nearly hydrostatic way, the cluster mass
can be determined from the intracluster plasma density and temperature
distribution through the hydrostatic equation, where density and
temperature profiles are determined from X-ray imaging and X-ray
spectroscopy (e.g.~Sarazin 1986). However, galaxy clusters are
dynamically young objects, that is, their dynamical timescale is not
much shorter than their age and the Hubble time. Therefore, in
contrast to elliptical galaxies for example, a significant fraction of
the clusters are not very close to a dynamical equilibrium. This
provides a challenge to mass determination. The validity and precision
of the mass determination with X-ray observations thus needs to be
tested.

Observations of the gravitational lensing effect of galaxy clusters
offer an alternative route to the determination of the cluster mass,
which is not dependent on the assumption of hydrostatic equilibrium of
the intracluster medium. This approach has other drawbacks, however,
since the mass determined refers to all the excess mass over the mean
density of the Universe in the line-of-sight of the cluster. Further
the gravitational lensing distortion signal statistics is limited by
the number of background galaxies that can be observed, which provides
a fundamental limit on the significance of the lensing analysis
results. Therefore large efforts are undertaken to provide an
inter-calibration of the two methods (Hoekstra 2007, Bardeau et
al. 2007, Pedersen \& Dahle 2007, Mahdavi et al. 2008, Leauthaud et
al. 2010). The Local Cluster Substructure Survey (LoCuSS) project
(Smith et al. 2005) involves one of the largest of these efforts
(Zhang et al. 2008, 2010, Okabe et al. 2010a,b) concentrating on
clusters at $0.15<z<0.3$.

In this context it is very important to investigate how possible
deviations in the mass measurement with the two methods depend on the
dynamical state of the galaxy clusters. For example, early
results from LoCuSS, based on characterising the X-ray morphology of
12 clusters using Asymmetry and Fluctuation parameters imply that
``undisturbed'' clusters are on average more massive than
``disturbed'' clusters in mass - $Y_X$ relations at $2\sigma$ significance 
(Okabe et al. 2010b). The $Y_X$ parameter is defined as the product of the 
gas mass and the temperature, which is believed to be a particularly good 
mass proxy (Kravtsov et al. 2006).

The main goal of this paper is, therefore, to measure
substructure in the X-ray surface brightness distribution of
clusters at $0.15<z<0.3$ as signatures of deviations from dynamical
equilibrium, and to compare with optical/near-infrared indicators of
cluster structure. We apply the currently most popular
substructure statistics of power ratios and center shifts to a sample
of 35 clusters which have deep X-ray observations with
\emph{XMM-Newton} in order to rank the clusters by their substructure in the
X-ray surface brightness image. These substructure parameters are then
compared to a number of physical cluster properties determined from
X-ray data. We also compare the occurrence of substructure with 
selected optical/near-infrared properties of the clusters, including
the luminosity gap between the first and second ranked cluster
galaxies, and the ellipticity of the brightest cluster galaxy. We
also investigate how cluster scaling relations vary as a function of
the degree of substructure. 

The paper is structured as follows. In section 2 we give a brief
description of the cluster sample, and outline the methods used in our
analysis in section 3. The results of the substructure analysis and
correlations with other cluster physical parameters are presented in
section 4, and section 5 provides a comparison of the substructure
measures with optical cluster parameters. The influence of
substructure on scaling relations is studied in section 6. The
conclusions and a summary are given in section 7.

Throughout the paper we adopt a flat $\Lambda$-cosmological model with
$\Omega_m$=0.3 and a Hubble constant of $H_0$=70~km/s/Mpc for the
calculations of all distant dependent quantities.

\section[]{Cluster sample and X-ray data reduction}

We select massive clusters from the X-ray galaxy cluster catalogues, 
BCS (Ebeling et al. 1998, 2000) and REFLEX (B\"ohringer et al. 2004), 
based on the ROSAT All-Sky Survey (Tr\"umper 1993) using the same 
selection criteria as LoCuSS: $L_X/E(z)^{2.7}\textgreater4.1\times10^{44}{\rm erg/s}$ 
in the 0.1--2.4keV band and z=0.15-0.3 with a line-of-sight interstellar 
column density $n_H<7\times10^{20}{\rm cm}^{−2}$ (Smith et al., in prep.; 
see also Marrone et al., 2012), with no restriction on declination.  
After cross-comparison with the \emph{XMM-Newton} archive
this yields a sample of 37 clusters with deep \emph{XMM-Newton} 
observations. This sample has previously been analysed by Zhang et
al. (2008). Here we analyse 35 of the 37 clusters of this
sample; the two clusters, Abell 115 and Abell 901 have not been
included here, because both are a part of larger structures. 
Abell 115 comprises two cores (e.g. Gutierrez \& Krawczynski, 2005, 
Barrena et al., 2007), and Abell 901 is a part
of a supercluster together with Abell 902 (e.g. Gray et al., 2002) 
where in both cases it is difficult to disentangle cluster emission 
from the emission of the rest of the environment.

The \emph{XMM-Newton} observations for all three detectors were flare-cleaned, 
and out-of-time events were statistically subtracted from the pn data. Point 
sources and other sources unrelated to the galaxy clusters were removed 
and the holes were statistically filled with photons from surrounding regions. 
The images from all three detectors were combined and the corresponding 
exposure maps were added with an appropriate weighting for the MOS detectors 
to bring them to the level of an effective pn exposure. The substructure 
analysis was then applied to the exposure-corrected and background-subtracted 
combined images. In practice the analysis programs use the count, exposure and
background maps simultaneously to reconstruct all relevant quantities related 
to photon noise. The \emph{XMM-Newton} data and substructure analysis
is described in more detail in B\"ohringer et al. (2010).

\section[]{Characterising substructure}

To characterise the degree of substructure of clusters in our sample,
we concentrate here on two methods: centre shifts as a function of the
aperture radius (e.g. Mohr et al. 1993, 1995; Poole et al. 2006,
O'Hara et al. 2006) and the determination of so-called power ratios
(e.g. Buote \& Tsai 1995, 1996; Jeltema et al. 2005, 2008, Valdarnini
2006, Ventimiglia et al. 2008, B\"ohringer et al. 2010). We use the
same technique as in B\"ohringer et al. (2010) with some modifications
as outlined below.

\subsection[]{Power ratios}

The power ratio method was introduced by Buote \& Tsai (1995) using
the X-ray surface brightness as a representation of the projected mass
distribution of a cluster. A multipole decomposition of the potential of 
the two-dimensional, projected mass distribution then provides moments 
which are identified as power ratios once they are normalised by the zeroth 
moment. In the analysis of Buote \& Tsai, the moments, $P_m$ are determined 
as follows

\begin{equation}
P_0 = \left[ a_0 \ln (R_{\rm ap}) \right]^2
\end{equation}

\begin{equation}
P_m = { 1 \over 2 m^2 R_{\rm ap}^{2m} } \left( a_m^2 + b_m^2 \right)
\end{equation}

\noindent where $R_{\rm ap}$ is the aperture radius. The moments $a_m$
and $b_m$ are calculated by:

\begin{equation}
a_m(r) = \int_{r \le R_{\rm ap}} d\vec{x} ~S(\vec{x}) ~r^m \cos (m\phi)
\end{equation}

\noindent and

\begin{equation}
b_m(r) = \int_{r \le R_{\rm ap}} d\vec{x} ~S(\vec{x}) ~r^m \sin (m\phi), 
\end{equation}

\noindent where $S(\vec{x})$ is the X-ray surface brightness image,
and the integral extends over all pixels inside the aperture radius.
$a_0$ in Eq. (1) is thus the total radiation intensity inside the
aperture radius.

Since all terms, $P_m$, are proportional to the total intensity of the
cluster X-ray emission, while only the relative contribution of the
higher moments to the total emission is of interest, the moments are 
normalised by $P_0$, resulting in the so-called power ratios, $P_m/P_0$. 
Similarly to all previous studies, we only make use of the lowest moments 
from $P_2$ to $P_4$ (quadrupole, hexapole and octopole). Before the multipole 
moments are determined, the center for the calculations is found by determining 
the photon count center within the aperture used. The dipole, $P_1$,
should therefore vanish, which is checked during the calculations.
$P_2/P_0$ describes the quadrupole of the surface brightness
distribution, and is not necessarily a measure of substructure; a
quadrupole will also be detected for a very regular elliptical cluster. 
In practice, low to moderate values of $P_2/P_0$ are found
for regular elliptical clusters, while larger values of $P_2/P_0$ are
a sign of cluster mergers. The lowest power ratio moment providing a 
clear substructure measure is thus $P_3/P_0$. $P_4/P_0$ describes
substructure on slightly finer scales and is found to be correlated with 
$P_2/P_0$ here and in previous studies.

The uncertainty of the power ratio measurement and the influence of
photon noise are studied by simulations in which an additional
Poisson-noise is imposed on the count images with background. This
``Poissonisation'' is equivalent to the Poisson-noise introduced by
the observation involving a finite number of detected photons. We
interpret the variance of the power ratio results from the simulations
as the measurement uncertainty. We also investigate how much
additional signal is introduced by the artificial Poissonisation. We
then make the assumption that the additional power introduced by the
Poissonisation is similar to the extra power in the power ratios
introduced by the photon shot noise of the observation. We therefore
subtract the additional noise found in the mean of all simulations
compared to the observations from the observational result.

Power ratios are routinely determined in 10 different apertures with
radius $r \le n\times0.1 r_{500}$ (with n=1,2,$\ldots$,10). Different
to the prime approach in B\"ohringer et al. (2010) we are not excising
the cluster core in the surface brightness distribution in our
substructure analysis. A core excision has been found to have not much
influence on the result except for the smallest aperture.

\subsection[]{Modified power ratio calculation}

We also introduce a couple of modifications to the calculation of power 
ratios motivated by the following arguments. The power ratios evaluated at 
$r_{500}$ with the $r^{2m}$ weighting are predominantly influenced by the 
dynamical state in the outskirts of a cluster. In turn this means that 
disturbances near the centre of the cluster, especially if their scale is 
relatively small compared to $r_{500}$, will be severely weighted down. 
To overcome this strong bias to the outer part of the cluster, we measure 
the averaged power ratios at ten radii defined by fractions of $r_{500}$.

An alternative approach to reduce the strong weighting of the outer region 
uses a different weighting in $r$. We have chosen a linear weighting in $r$ 
(instead of $r^m$ in Eqs. (3) and (4)), since for an approximate radial 
dependence of the surface brightness proportional to $r^{-2}$ this procedure 
gives similar weight to linearly spaced annuli. We expect these two modified 
approaches to yield similar results.

\subsection{Centre shifts}

In most recent work the centre shift method is based on a measurement of the 
variance of the separation between the X-ray peak and the centroid calculated 
within concentric apertures with increasing aperture size. Instead of 
comparing the aperture centroids with the separately determined X-ray maximum, 
we measure the centroid shift for each aperture with respect to the mean of 
all centroids. We will explain the reason for this modification below. The 
centroid of each aperture is found by determining the ``center of mass'' of 
the photon distribution in each aperture, which was already used for the 
centering prior to the power ratio determination. The resulting $w$ is then 
the standard deviation of the different centre shifts (in units of $r_{500}$), 
defined as (see also Poole et al. 2006):

\begin{equation}
w~ =~  \left[ {1 \over N-1}~ \sum \left( \Delta_i -  \langle 
\Delta \rangle \right)^2 \right] ^{1/2} ~\times~ {1 \over r_{500} } 
\end{equation}

\noindent where $\Delta_i$ is the distance between the mean centroid
and the centroid of the $i$th aperture. We have decided to use the
mean centroid value of all apertures as the reference centre. One
finds that the variance, $w$, of the centre shift determined with
Eq. (5) depends on the choice of the reference centre. Rather than
using a reference centre determined in a completely different way - as
in the case where the X-ray maximum is adopted for the reference
centre - we prefer a reference centre obtained within our system of
centre determinations. A logical choice is then to take the mean centre
of all apertures as reference. Comparing this approach to the previously 
used X-ray maximum reference, we find that the new method yields on
average slightly smaller $w$ values. Consequently we find as described
below smaller values of $w$ for convenient boundaries between regular
and disturbed clusters.

The uncertainties in the centre shifts and in the $w$ parameter 
are determined with the same simulations as the uncertainties 
of the power ratios, i.e., by using Poissonised re-sampled 
cluster X-ray images. The standard deviation of the $w$ parameter 
in the simulation is used as an estimate of the measurement 
uncertainties. We do not use the noise-bias subtracted $w$-parameter 
as in the case of the power ratios since the bias correction is
mostly much smaller than the errors or the bias correction does not
shift the $w$-parameter to alter the classification of the cluster
morphology.

All error calculations for the power ratios and centre shifts 
properly include any correlations and systematics that may exist. 
This is ensured by the end-to-end Monte-Carlo simulations of the 
Poissonised data analysed exactly the same way as we calculate 
the power ratios and centre shifts with the original X-ray data.

\section[]{Results}

In the following we will use the average value of $P_3/P_0$ from all rings 
and the new definition for the $w$-parameter as our primary results for the 
discussion. The power ratios reported in the following are determined without 
core excision. All biases and errors resulted from one thousand simulations. 
Where power ratios fall below zero, we kept them at the left-most side of 
each plot truncating their lower error bars. Morphological parameters for 
individual clusters are listed in Table~\ref{tab:app} in the Appendix. 
Clusters are numerically labelled in descending order by the centre 
shift values in the legend of all figures. For convenience we abbreviate 
$P_m / P_0$ as $P_m$ for the rest of this paper except the figure axis labels. 
Throughout this paper we used $r_{500}$ and $M_{500}$ determined in Zhang et 
al. (2008) for the consistency.

\subsection[]{Comparing Power Ratios and Centre Shifts}

$P_3$ is the first moment that contains information on deviations from 
azimuthal symmetry above a simple ellipticity. We therefore explore the 
$w-P_3$ plane first, and subsequently inspect the $P_2 - P_3$ and $P_4 - P_3$ 
spaces to further investigate aspects of X-ray cluster morphology.

\begin{figure}
\begin{center}
\resizebox{\hsize}{!}{\includegraphics[height=6cm]{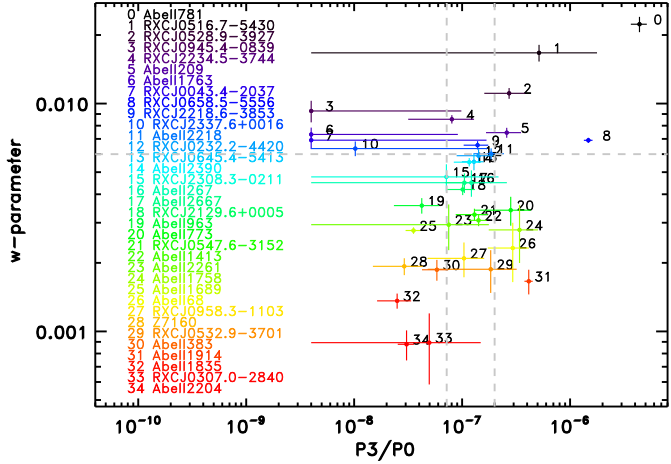}}
\end{center}
\caption{Centre shift and averaged $P_3$ parameters.
  Clearly seen are the five most disturbed clusters in the upper right
  corner with the largest $w$ and $P_3$ values. The intermediate cases
  are bounded by the two vertical dashed lines below the horizontal
  line, and the regular clusters are located in the left lower corner.
  Clusters 3, 6, 7, and 10 have very low $P_3$ values, however, they
  are classified as disturbed given the large centre shifts. In the
  same way 20, 24, 26, and 31 are disturbed due to their large $P_3$
  values. Clusters are numerically labelled by their $w$ values in
  descending order.}
\label{fig:w-p3}
\end{figure}

Fig.~\ref{fig:w-p3} shows the $w$-parameter against the modified and
normalised $P_3$ for the sample as described in the previous section.
The distribution of the substructure parameters is continuous as is
the morphological appearance of clusters and there are no very obvious 
criteria to draw boundaries between different sub-classes of clusters. 
Indeed, detailed inspection of sensitive observations of any cluster reveals 
some asymmetric features, therefore it is problematic to classify a subset of
clusters as (for example) completely ``relaxed''. Nevertheless, to aide the 
discussion that follows, we sub-divide the clusters in our sample into three 
classes based on their locations in the $w-P_3$ plane, as delineated by three 
gray dashed lines. First we draw attention to the five clusters in the upper 
right corner that have the largest centre shifts and $P_3$ -- these five 
include famous merging clusters such as the ``Bullet cluster'', which is 
number 8 in our sample. However we use wider criteria to classify clusters as
``disturbed'': $w\textgreater0.006$ or $P_3\textgreater2\times10^{-7}$, i.e.\ above the horizontal 
gray line and/or to the right of the right vertical gray line. Clusters in 
the lower left rectangle display the smallest centre shifts and lowest $P_3$ 
-- we classify clusters that satisfy $w<0.006$ and $P_3<7\times10^{-8}$ as 
having a ``regular'' X-ray morphology. Finally, we classify clusters in the 
lower narrow rectangle surrounded by the gray dashed lines as having an 
``intermediate'' X-ray morphology -- i.e.\ $w<0.006$ and 
$7\times10^{-8}<P_3<2\times10^{-7}$. The results on the classification are 
summarised in Table~\ref{tab:app} of the Appendix. In summary we classify 8 
clusters as regular, 11 clusters as intermediate, and 16 clusters as disturbed.
We stress that this classification is not unique and clusters have a rather 
continuous distribution of morphologies. We have visually checked this
classification scheme, and found it to give the most consistent picture 
in a comparison of the visual impression and the numerical classification 
compared to the studies described below.

\begin{figure}
\begin{center}
\resizebox{\hsize}{!}{\includegraphics[height=6cm]{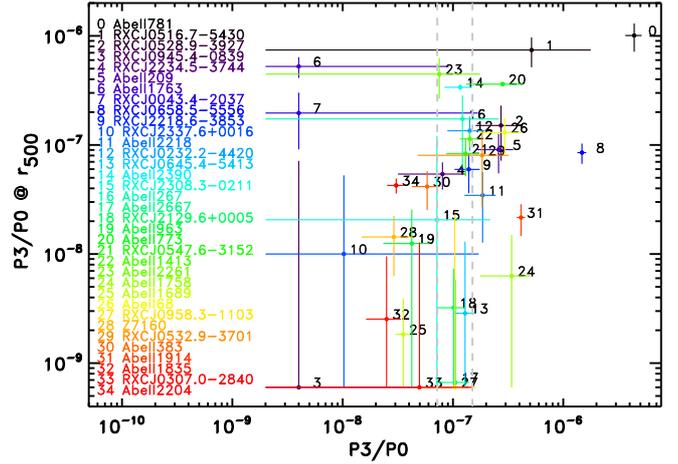}}
\end{center}
\caption{Comparison of $P_3$ calculated at $r_{500}$, and averaged over
  the ten rings. A general linear correlation can be seen with some scatter as
  expected. These two ways of calculating $P_3$ values classify a number of 
  clusters differently, see text for details. 
}
\label{fig:p3-p3r500}
\end{figure}

To set this result using the radially averaged $P_3$ value into perspective 
with previous results, we compare in Fig. 2 the average $P_3$ result with the 
$P_3$ values determined for the largest aperture, $r_{500}$. We expect these 
values to be almost linearly correlated, except for those cases in which 
$P_{3,r500}$ boosts a structure near the aperture radius or overlooks a 
structure in the central region. Indeed clusters 16, 21, 23, 29, 30, and 34 
would have been identified as disturbed, and 13, 17, 24, and 27 as more regular
if the traditional $P_3$ value was adopted. On the other hand clusters 6 and 7 
have much larger $P_3$ evaluated at $r_{500}$. They would have been identified 
as regular clusters based solely on the averaged $P_3$, however, as seen in 
Fig. 1, our joint $w-P_3$ classification identifies these clusters as 
disturbed systems.

\begin{figure}
\begin{center}
\resizebox{\hsize}{!}{\includegraphics[height=6cm]{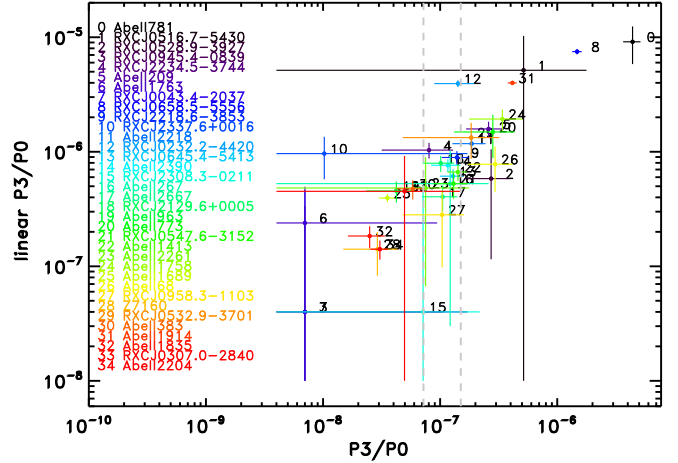}}
\end{center}
\caption{Comparison of $P_3$ parameters calculated with a linear weight
  in the radial direction to the averaged $P_3$ over the ten rings. A similar
  trend is shown as in Fig. 2 with a tighter scatter. It shows that
  our modified $P_3$ calculation is closer to a linear weighting in
  radius instead of $r^{m}$.}
\label{fig:p3-p3linr}
\end{figure}

We also tested another approach to determine the hexapole moment $P_3$. 
Assuming the surface brightness profile of a cluster falls with $r^{-2}$ 
where the beta parameter is close to 0.5, and the area of a ring scales as 
$\Delta * r$, where $\Delta$ is a fixed increment such as 0.1 to 1, we find 
that a scaling proportional to $r$ cancels both effects. Specifically $P_3$ 
from the linear weighting in $r$ from Eqs. (2)--(4) is shown in Fig. 3 
compared to the average $P_3$. One notes a clear linear correlation except 
cluster Nos. 6, 10, and 15, which, however, accompany very large 
uncertainties. $P_3$ calculated by the traditional method represents mainly 
the asymmetries in the outskirts of clusters while the dynamical state near 
the core is almost ignored. A cluster with a merging event localised at the 
centre, for example, will not be identified as disturbed if the traditional 
method is used. In our new approach the averaged $P_3$ can trace closely any 
distinct changes in the dynamical state in a nearly equal weight in $r$ as 
seen in Fig. 3. Note that on average the absolute values of the different 
definitions of $P_3$ are expected to change.

\begin{figure}
\begin{center}
\resizebox{\hsize}{!}{\includegraphics[height=6cm]{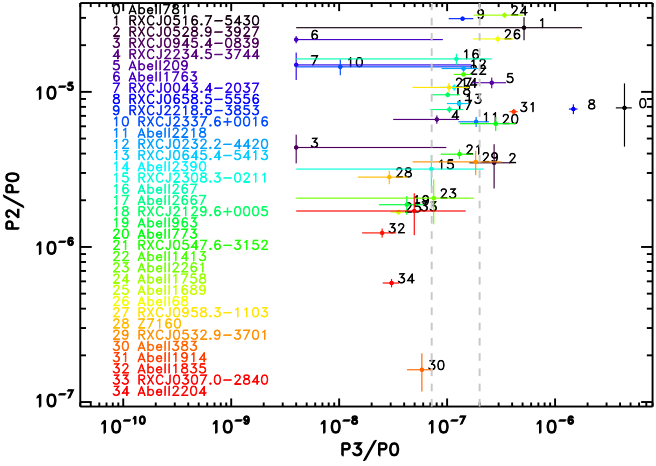}}
\end{center}
\caption{
  Comparison of the $P_2$ and $P_3$ parameters. Easily seen are the 
  three disturbed clusters, 6, 7, and 10 in the top left corner flagged 
  by large $P_2$ values.
}
\label{fig:p2-p3}
\end{figure}
\begin{figure}
\begin{center}
\resizebox{\hsize}{!}{\includegraphics[height=6cm]{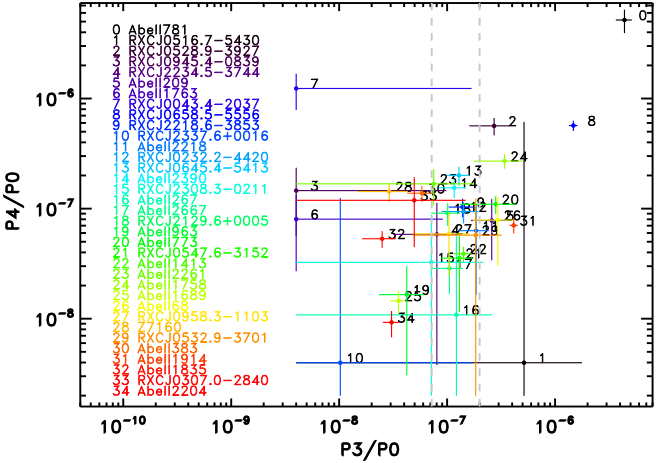}}
\end{center}
\caption{
  Similarly $P_4$ and $P_3$ are compared. Cluster 7 is a clearly disturbed
  system as given by its $P_4$ value although it has a a very low $P_3$.
}
\label{fig:p4-p3}
\end{figure}

Because the hexapole moments, $P_3$, may not catch all forms of disturbed 
morphologies, as for example very elongated or bimodal merging clusters 
which are nearly axisymmetric, we also inspect the results of the quadrupole 
and octopole analysis $P_2$ and $P_4$. Fig. 4 shows a diagram of the 
distribution of $P_2$ and $P_3$ values. While there is a good correlation of 
two parameters, we immediately recognise two outliers with low $P_3$ and high 
$P_2$ values, the cluster Nos. 6 and 7. Both are elongated merging clusters
and both have already been flagged as disturbed by their high centre shift 
values, $w$. Cluster 10 is a similar less extreme case also flagged as 
disturbed by its $w$ value.

Fig. 5, which presents the distribution of $P_3$ and $P_4$ values shows again 
the clusters 6 and 7 with high $P_4$ and low $P_3$ values. Another cluster 
with a relatively high $P_4$ value is cluster no. 3, again an object with a 
high $w$-parameter in Fig. 1.

\begin{figure}
\begin{center}
\resizebox{\hsize}{!}{\includegraphics[height=6cm]{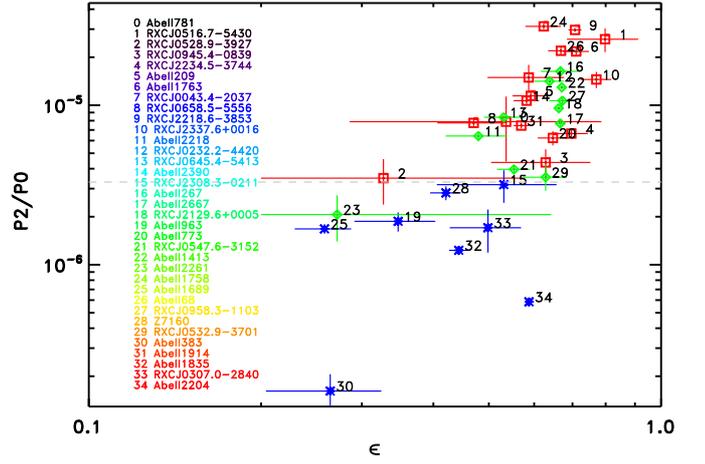}}
\end{center}
\caption{Comparison of $P_2$ and the ellipticity, $\epsilon$, 
  calculated from a fit of an elliptical $\beta$ model to the X-ray image. 
  The blue stars represent those classified as regular, the green diamonds 
  the intermediate, and the disturbed clusters are marked by the red squares. 
  This convention is carried throughout the paper. 
  There is a positive correlation between two parameters. We note that there 
  is a segregation effect which separates regular clusters from the 
  intermediate and disturbed ones.}
\label{fig:p2-ellip}
\end{figure}

In summary the inspection of the $P_2$ and $P_4$ parameters did not change 
our classification, rather reinforced our previous finding. To further explore 
the meaning of the $P_2$ parameter, we also determined the ellipticity of the 
clusters in the X-ray images of the sample directly, which is defined by 
$\epsilon = \sqrt{ (a^2-b^2)/a^2 }$ where $a$ is the semi-major and $b$ is the 
semi-minor axis of the fitted elliptical $\beta$ surface brightness model. 
Fig. 6 shows that there is a good correlation of $\epsilon$ and $P_2$. 
According to the three morphological classes the clusters are colour 
(symbol)-coded as shown in Fig. 6. The blue stars are those classified as 
regular, the green diamonds represent the intermediate, and the disturbed 
clusters are marked by the red squares. This convention is carried through for
the rest of the paper. We see a clear segregation of all regular clusters and 
one intermediate cluster having $P_2$ values below $\approx 3\times 10^{-6}$ and 
the disturbed and all other intermediate clusters lying above this devision. 
For the ellipticity, $\epsilon$, the trend is also there but the segregation 
is less strong. Fig.~\ref{fig:p2-ellip} seems to give the strong implication 
that the quadrupole moments up to about $P_2\approx $ $3\times 10^{-6}$ refer 
mostly to nearly relaxed elliptical clusters, while larger values of $P_2$ 
are caused by dynamical distortions of the clusters.

In Table~\ref{tab:corr} we list a number of the correlation 
coefficients for the most relevant cases in our paper with
the Kendall and Spearman tests. We use the ASURV (Astronomical
Survival Statistics, Isobe et al. 1986) package to calculate
the correlations in the presence of censored data. The correlation 
for the $w$-$P_3$ pair gives probabilities of 0.19 for no correlation 
according to Kendall's $\tau$, and 0.28 to Spearman's $\rho$ test. 
When we re-calculate this correlation excluding the clusters 3, 6, 
and 7 in Fig. 1, we obtain much 
smaller probabilities for no correlation, 0.006 for Kendall's and
0.01 for Spearman's tests giving stronger significance to the 
correlation. However, we emphasise that the morphological 
classification that we adopt in this paper relies on both $w$ and 
$P_3$ parameters as explained above, and the correlation found
between these two parameters is not necessarily expected to be 
generalised without exceptions as demonstrated by clusters 3, 6, and 7. 

The correlation coefficient for $P_2$ and the X-ray ellipticity,
$\epsilon$ is also shown in Table~\ref{tab:corr}. It is one of the most 
tightly correlated relations confirming quantitatively our visual 
impression in Fig. 6. This relation gives vanishing probability for no 
correlation for both statistical tests. 

\begin{table}
  \caption{{\footnotesize Correlation coefficients measured for various 
    cluster and substructure parameters. 
    (1) the correlation tested 
    (2) the coefficients of Kendall's $\tau$ test 
    (3) the corresponding probability of a null-correlation 
    (4) the coefficients of Spearman's rank test, $\rho$ 
    (5) the corresponding null probability 
    (6) the corresponding figure number that shows the correlation. 
    We used the ASURV software package to calculate the coefficients.
    }
  }
  \label{tab:corr}
  \begin{center}
  
  \begin{tabular}{l r r r r r}
    \hline
    \hline
    \\
    \multicolumn{1}{l}{Correlation} & 
    \multicolumn{1}{c}{$\tau$} & 
    \multicolumn{1}{c}{$P_{\mathrm null}$} & \multicolumn{1}{c}{$\rho$} &
    \multicolumn{1}{c}{$P_{\mathrm null}$} & \multicolumn{1}{c}{Figure} \\
    (1)  &  (2)  &   (3)  &    (4)   &  (5)  & (6)  \\
    \hline
    \\
    $w-P_3$     & 1.31 & 0.19 & 0.18 & 0.28 & 1\\
    $w-P_3^*$   & 2.74 & 0.006 & 0.41 & 0.01 & 1\\
    $P_2-\epsilon$              & 4.70 & 0.00 & 0.74 & 0.00 & 6\\
    $w-$luminosity ratio        & 4.87 & 0.00 & -0.76 & 0.00 & 8\\
    $P_3-$luminosity ratio      & 2.46 & 0.01 & -0.36 & 0.03 & 8\\
    $P_3-$luminosity ratio$^*$  & 2.97 & 0.003 & -0.44 & 0.01 & 8\\
    luminosity gap-$P_3$   & 1.10 & 0.26 & -0.19 & 0.35 & 9 \\ 
    luminosity gap-$P_3^*$ & 2.25 & 0.02 & -0.41 & 0.04 & 9 \\ 
    BCG $\epsilon-P_2$          & 2.77 & 0.006 & 0.66 & 0.007 & 10\\
    BCG $\epsilon-$X-ray $\epsilon$ & 2.39 & 0.01 & 0.63 & 0.009 & 10\\
    \\
    \hline
  \end{tabular}
\end{center}
Note: $^*$ denotes the same correlation as in the preceding row  
but excluding the three clusters, 3, 6, and 7 as for example 
shown in Fig. 1. For the correlation of the luminosity gap and $P_3$ 
the corresponding clusters, 2, 4, and 5 in Fig. 9, are excluded.
\end{table}

\subsection[]{Correlation of cluster cool-core parameters and
  morphology}

We now compare our ranking of clusters based on measured substructure 
parameters with physical cluster parameters. First, we consider the existence 
of a cool core in the centres of clusters signified by a strong central 
density increase and a central temperature drop (e.g. Peterson \& Fabian 2006, 
Hudson et al. 2010). In the X-ray surface brightness distribution a cool
core shows up as a bright central cusp. Hence the ratio of the
luminosity with and without the core can be used to divide the cluster
sample into cool and non-cool core clusters.

\begin{figure}
\begin{center}
\resizebox{\hsize}{!}{\includegraphics[height=6cm]{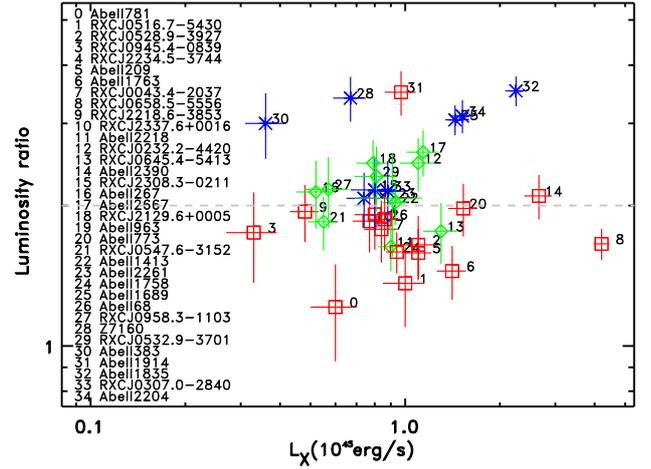}}
\end{center}
\caption{
  Luminosity ratio as a function of the core-excised luminosity. 
  A clear trend is shown with the horizontal line, 
  which marks the median of the luminosity ratio where the regular clusters 
  have a higher luminosity ratio compared to the disturbed.
}
\label{fig:lratio-lx}
\end{figure}

We take the core-excised and core-included X-ray bolometric luminosity of the 
clusters from Zhang et al. (2008) where the core was defined by 
$r \textless$0.2$r_{500}$. Fig.~\ref{fig:lratio-lx} shows the X-ray luminosity 
ratio against the core-excised X-ray luminosity. For the rest of the paper 
the luminosity ratio refers to the ratio of two luminosities determined with 
and without the core, 0.2$r_{500}$. In addition, given the number statistics 
we group the regular and intermediate clusters together whenever a numerical 
evaluation is performed. The dashed line is the median of the luminosity ratio 
which divides the clusters roughly in two subsamples, the regular and 
disturbed systems with the intermediate in between. There is a clear trend 
that this division also separates regular and disturbed clusters. All regular 
clusters have large luminosity ratios, and the disturbed clusters have low 
values with two exceptions, Abell 1914 and Abell 2390. Abell 1914 falls 
somewhere in between intermediate and disturbed, and Abell 2390 is close to 
the boundary, but it is known to have a strong cool core 
(B\"ohringer et al. 1998).

\begin{figure}
\begin{center}
\resizebox{\hsize}{!}{\includegraphics[height=6cm]{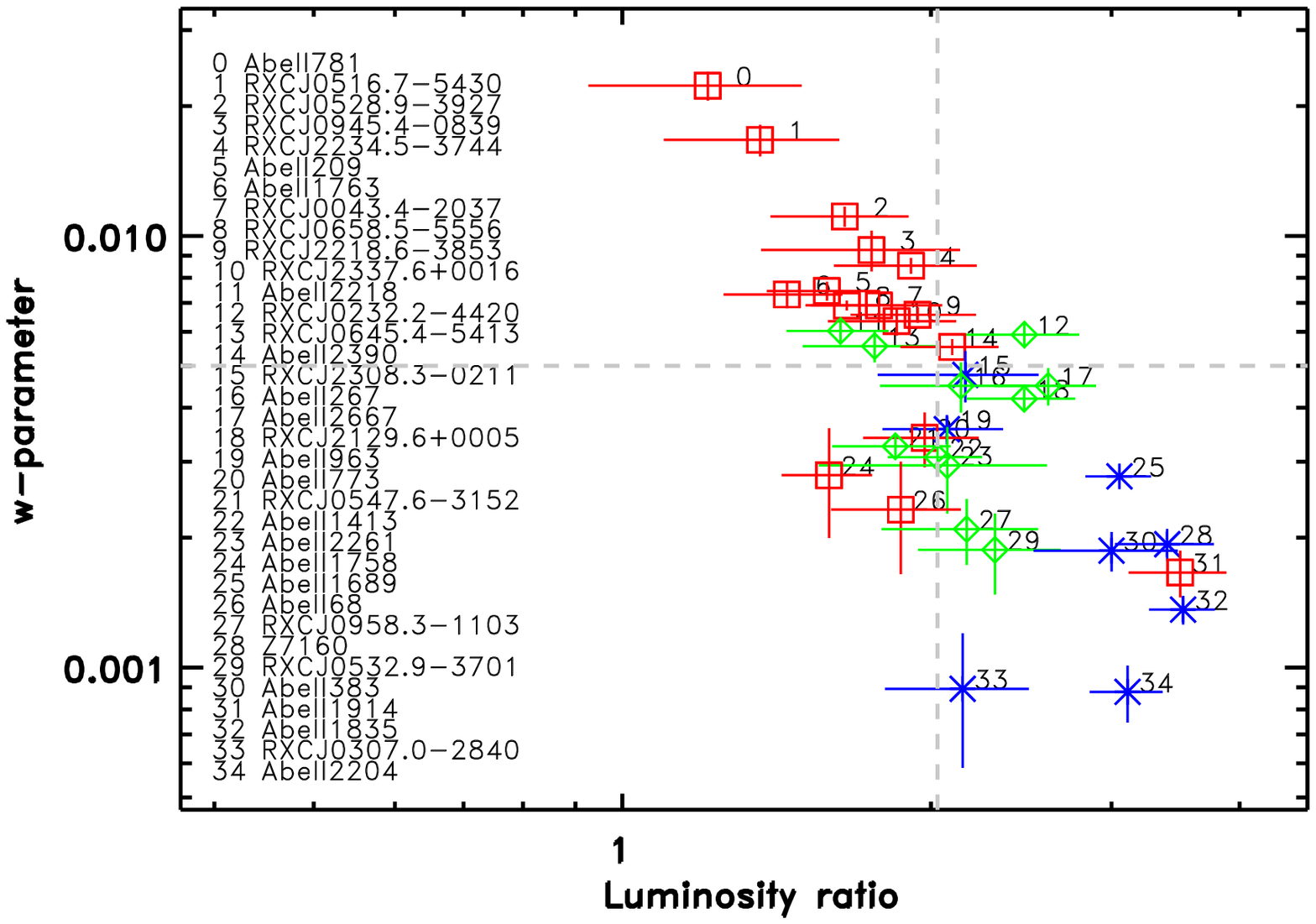}}
\resizebox{\hsize}{!}{\includegraphics[height=6cm]{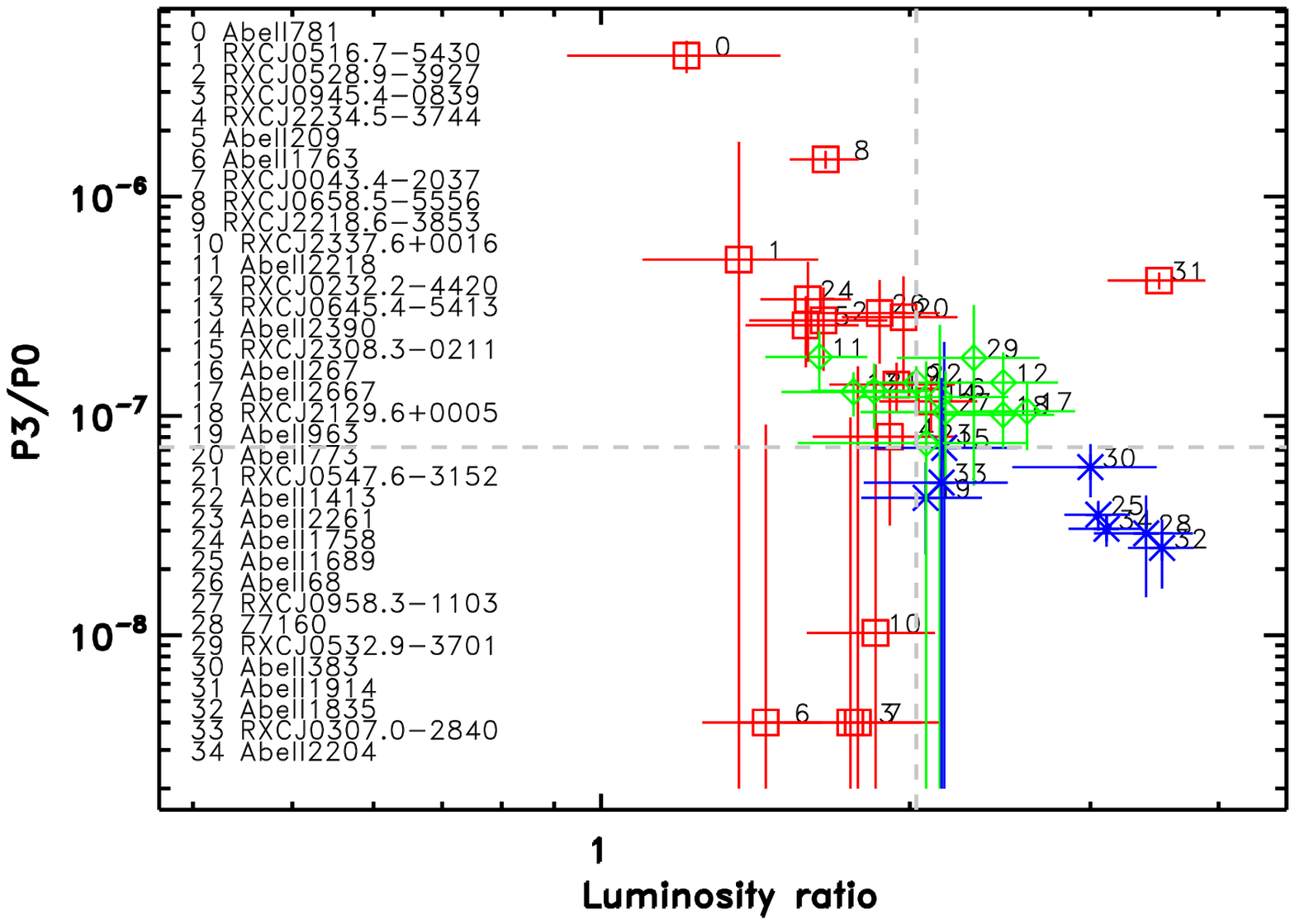}}
\end{center}
\caption{
  Structural parameters as a function of the luminosity ratio
  where a high luminosity ratio is an indicator of a cool core in the
  cluster. The regular clusters have the lowest $w$ values, and the
  largest luminosity ratio as shown in the upper panel. This trend is
  repeated for $P_3$ in the lower panel.
}
\label{fig:w-p3-lratio}
\end{figure}

Fig.~\ref{fig:w-p3-lratio} shows the correlations between the two structural 
parameters and the luminosity ratio. The upper panel shows the $w$-parameter 
against the luminosity ratio. Two dashed lines, which represent the median of 
the luminosity ratios and $w$ values from Fig. 1, divide the clusters in two 
groups. Notably the regular clusters are confined in the quadrant with the 
largest luminosity ratios and smaller $w$. While the intermediate clusters 
are spread in between, two clusters, Abell 2390 and Abell 1914 are the 
exceptions to be cool-core clusters flagged as disturbed by the $P_3$ parameter.
A similar trend is shown in the lower panel of Fig.~\ref{fig:w-p3-lratio} 
for $P_3$. Here four other clusters appear as exceptions, 3, 6, 7, and 10, 
which are not recognised as disturbed by $P_3$ but are flagged as such by 
their $w$-parameters. The correlations for both cases are quantified in 
Table~\ref{tab:corr}. $w$ and luminosity ratio are tightly anti-correlated 
with zero probability of being uncorrelated. Its Kendall's $\tau$ is 4.87, 
and Spearman's $\rho$ is -0.76. A similar anti-correlation is seen for the 
$P_3$ and luminosity ratio pair with the probability of 0.01 and 0.03 to be 
uncorrelated. This null-probability shrinks by a factor of three if the three 
clusters, 3, 6, and 7 are excluded. Finally, we note that we obtain similar 
results if we adopt the central electron density as our measure of the 
cool-core strength instead of the luminosity ratio.

\section[]{Correlations with Optical/near-infrared Parameters}

\subsection[]{Comparison of Luminosity Gap with X-ray Morphology}

We now consider the so-called luminosity gap -- i.e.\ the difference between 
the apparent magnitude of the brightest cluster galaxy (BCG) and the second 
brightest cluster galaxy, denoted by $\Delta m_{12}=m_1-m_2$. 
Smith et al.\ (2010) showed that clusters with dominant BCGs 
($\Delta m_{12}\textgreater1$) form a relatively homogeneous population of concentrated 
cool-core clusters with negligible substructure and an elliptical or disky 
BCG morphology. In contrast clusters with $\Delta m_{12}<1$ form a 
heterogeneous population. In summary, Smith et al.\ interpreted their results 
as indicating that luminosity gap probes a combination of both the formation 
epoch and recent assembly history (i.e.\ last few Gyr -- roughly one crossing
timescale) of a cluster. Specifically, the homogeneity of clusters with 
dominant BCGs suggests that such clusters formed relatively early and have 
not suffered significant in-fall recently. Such clusters would be expected 
to have relatively undisturbed X-ray morphologies, likely residing in the 
regular category discussed above.

\begin{figure}
\begin{center}
\resizebox{\hsize}{!}{\includegraphics[height=6cm]{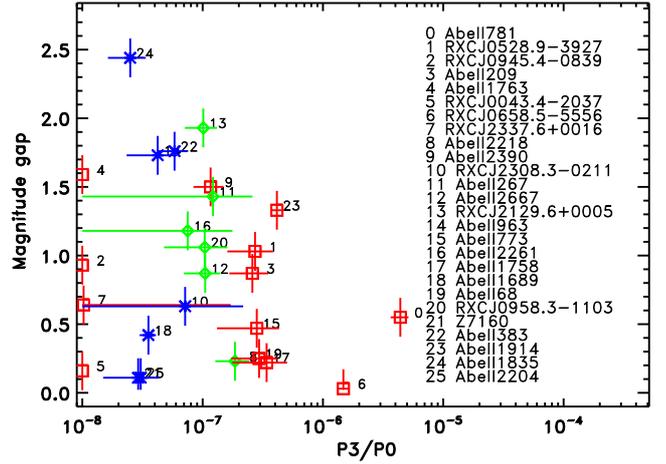}}
\end{center}
\caption{
  Luminosity gap compared to $P_3$. In total there are 26
  clusters whose luminosity gap data is available. Clusters with
  large luminosity gaps are a relatively homogeneous population of
  clusters with regular/intermediate X-ray morphology, with clusters
  that host less dominant BCGs presenting a wider range of X-ray
  morphology including the most extreme disturbed morphologies. 
}
\label{fig:gap-p3}
\end{figure}

To test Smith et al.'s interpretation we compare the X-ray 
morphologies that we measure in this article with luminosity gap
measurements for the 26 clusters in common between the respective
samples. Fig.~\ref{fig:gap-p3} shows that the four clusters with
the most dominant BCGs ($\Delta m_{12}\textgreater1.7$) have exclusively low
hexapole moments ($P_3\le10^{-7}$) and are classified exclusively as
having regular/intermediate X-ray morphology. At less extreme
luminosity gap the picture is less clear-cut, with clusters at
$\Delta m_{12}\textgreater1$ including a few disturbed systems, albeit at
$P_3\sim10^{-7}-4\times10^{-7}$. Clusters with low luminosity gaps
($\Delta m_{12}\textless1$) have the full range of hexapole moments seen in
the present sample. These results are consistent with Smith et
al.'s interpretation, although some of the detailed differences
(e.g.\ the disturbed X-ray morphology of some clusters with $\Delta
m_{12}\textgreater1$) is likely due to the differences between the apertures
within which the respective measurements are made. Specifically,
the field of view of Smith et al.'s data necessitated measuring
$\Delta m_{12}$ within a fixed physical radius of $640{\rm kpc}$ at
each cluster redshift, in contrast to the mean measurements of $P_3$
plotted in Fig.~\ref{fig:gap-p3}. 
To quantify this seemingly weak correlation we calculated the correlation 
coefficients in Table~\ref{tab:corr}. Kendall's $\tau$ coefficient is
1.1 with the null probability of 0.26 and Spearman's $\rho$ test gives 
-0.19 with 0.35 when including all data points. The correlation becomes
much tighter without those three corresponding clusters, which were excluded 
for other correlation calculations. The null probability for Kendall's $\tau$ 
is ten times smaller, 0.02 while that for Spearman's $\rho$ test is 0.04. 
A detailed investigation of optical
and X-ray cluster morphology will be presented in a separate article.

\subsection[]{Comparison of BCG Ellipticity with X-ray Morphology}

Previous studies have shown that the orientation of most
BCGs is roughly aligned with the elongation of the cluster X-ray
emission, and thus with the elongation of the cluster mass
distribution under the assumption that clusters are in approximate
hydrostatic equilibrium (e.g. B\"ohringer et al. 1997, Hashimoto et
al. 2008). 
The latter found a strong indication of an alignment between 
the cluster X-ray emission and BCGs, while no clear direct correlation 
between the ellipticity of X-ray and BCG was detected. Their sample 
includes a wide range of observational parameters in exposure times, 
redshift and fluxes, which may result in a much wider range of image quality. 
More recently Marrone et al. (2012) have pointed out
that the interpretation of BCG ellipticity is likely more subtle, in
that clusters hosting relatively circular ($\epsilon\textless0.2$) BCGs may
be elongated clusters that are observed along a line of sight close
to the major axis of an underlying prolate mass and gas
distribution. 

\begin{figure}
\begin{center}
\resizebox{\hsize}{!}{\includegraphics[height=6cm]{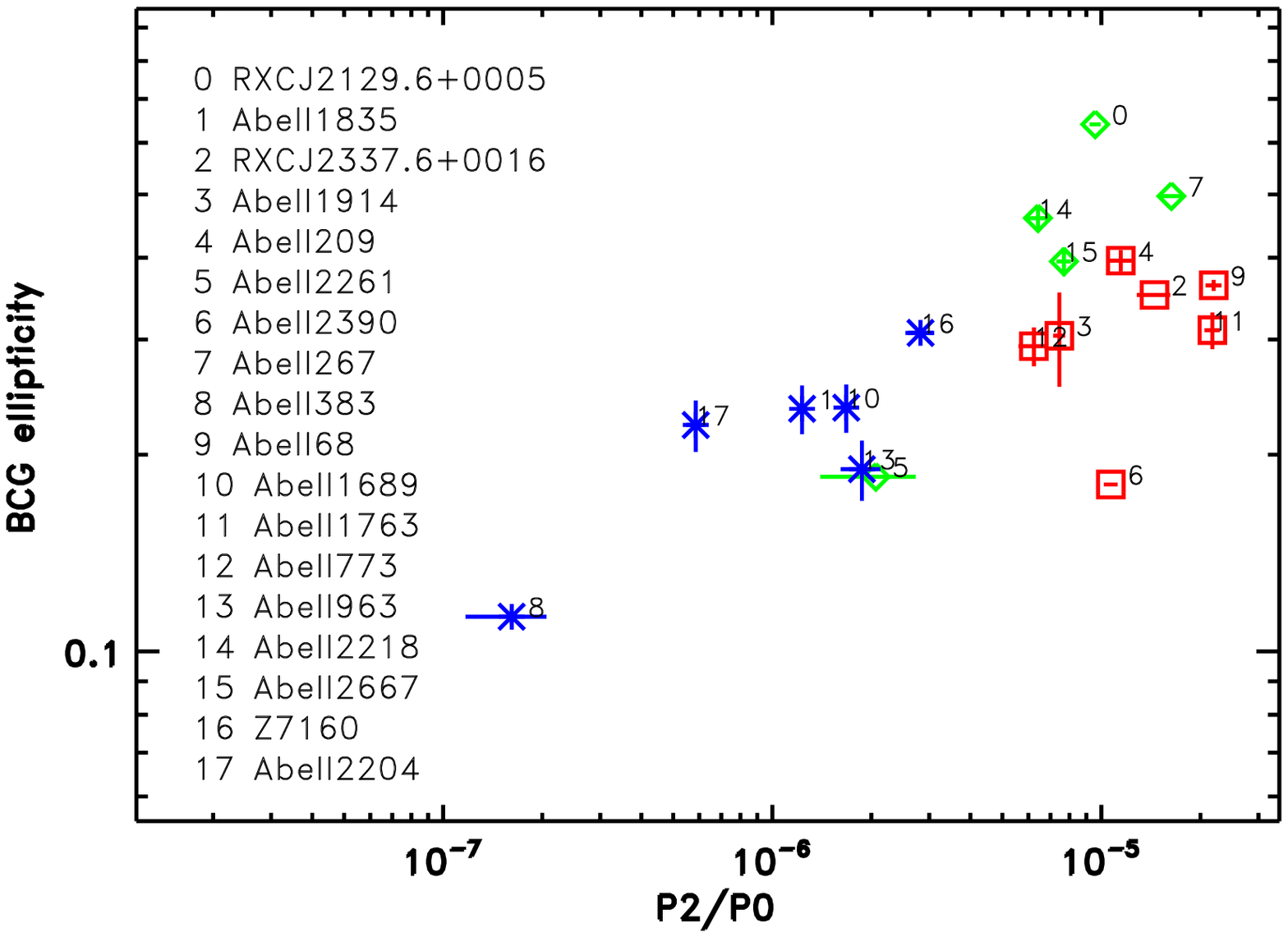}}
\resizebox{\hsize}{!}{\includegraphics[height=6cm]{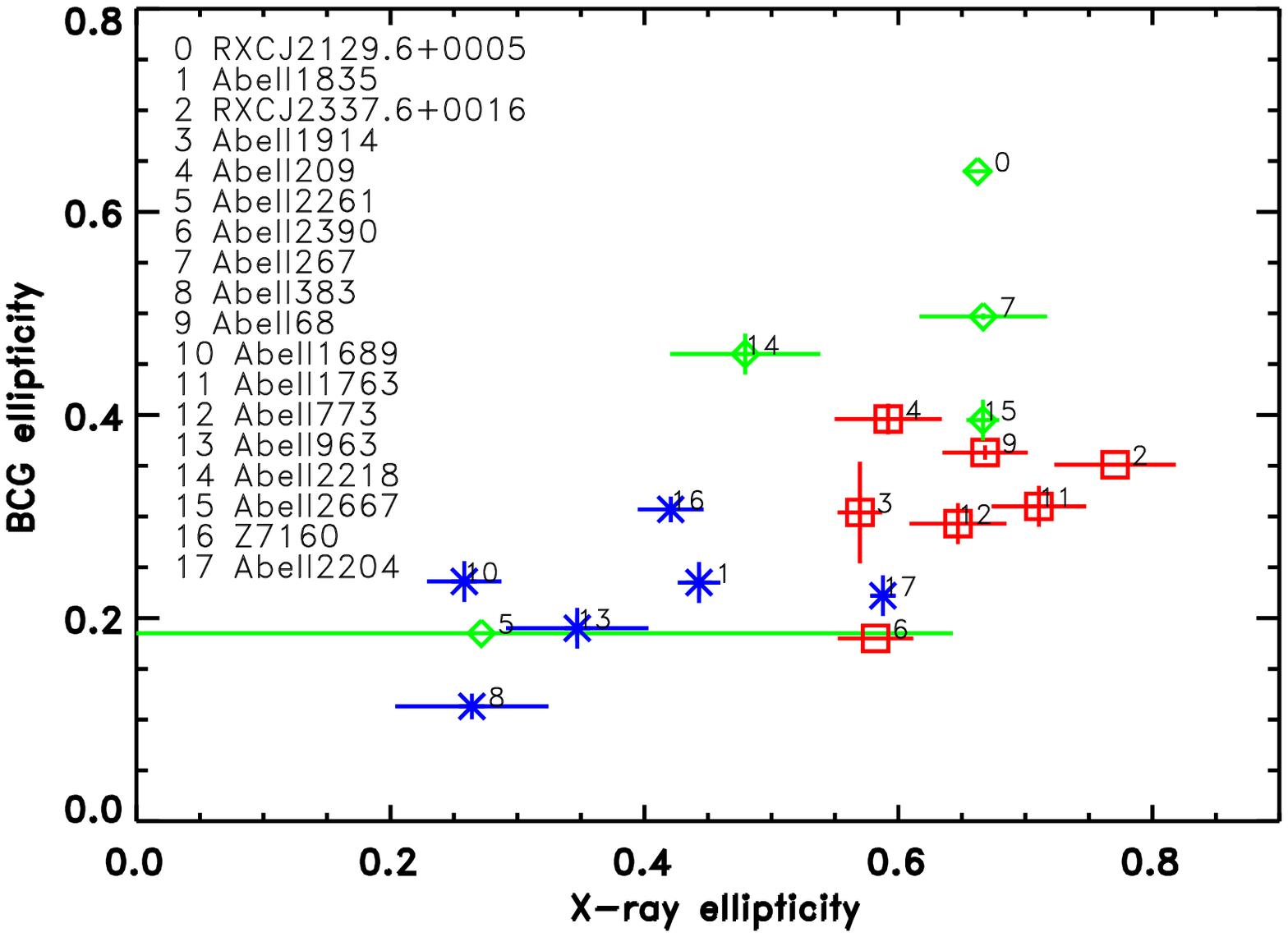}}
\end{center}
\caption{
  Comparison of the BCG ellipticity and the $P_2$ parameter
  for the clusters in the upper panel, and the cluster ellipticity in
  the lower panel. The cluster ellipticities were determined from the
  X-ray data as shown in Fig. 6, and the BCG ellipticities by
  Marrone et al. (2012).
}
\label{fig:ellip-ellip}
\end{figure}

In our study as shown in Fig.~\ref{fig:p2-ellip} we found a close
correlation of the ellipticity of the X-ray emitting gas and the
power ratio parameter, $P_2$. We therefore extend this comparison
by plotting in Fig.~\ref{fig:ellip-ellip} the ellipticity of the
BCG to the cluster $P_2$ parameter for a subsample of 18 clusters 
for which BCG ellipticity measurements are available (Marrone et
al. 2012). We also show in Fig.~\ref{fig:ellip-ellip} a comparison
of the BCG ellipticity with the X-ray ellipticity directly. 

In both comparisons we find a clear correlation with a larger
scatter for the cluster ellipticity measurements than for
the $P_2$ parameter. 
Quantitatively we show the correlations measured for both cases
in Table~\ref{tab:corr}. A Kendall's $\tau$ test gives a probability
of no correlation of 0.6\%, and similarly for a Spearman rank test
in the case of the BCG ellipticity and $P_2$ parameter.
The correlation is slightly decreased for the BCG ellipticity and
the X-ray ellipticity where the null-probability increases by a 
factor of 1.6 and 1.3, which still means 1\% of no correlation. 
Therefore an inspection of the upper panel
of Fig.~\ref{fig:ellip-ellip} is showing two effects:
BCGs with small ellipticities are living in clusters with small
elongations and also in the most regular clusters. Interestingly
the clusters hosting the BCGs with the largest ellipticities 
are classified as intermediate. These are the clusters with
moderately large $P_3$ values and small centre shifts. A 
closer look at the 4 clusters at the top of the BCG ellipticity
distributions shows that the host clusters are not
very much disturbed but rather elongated (elliptical). 
This raises the question what role line-of-sight projections
play in this effect. As suggested by Marrone et al. (2012),
BCGs and clusters with low ellipticities in projection 
could also imply prolate objects oriented with the major axis along
the line-of-sight. In this case the BCG or cluster ellipticity
could help to reveal the three-dimensional orientation. 
A detailed investigation of the degeneracy
between cluster orientation and the underlying structure of clusters
is beyond the scope of this article, and will be visited in a future
publication.

\section[]{Influence of substructure on scaling relations}

As outlined in the introduction, one of the main applications of the
substructure analysis is the study of the influence of substructure on
the scaling relations of X-ray observables and cluster mass. 
Previous studies of such scaling relations have used a variety of
methods to classify clusters, including the presence/absence of a
cool core, X-ray centroid shift, concentration and asymmetry
parameters, and offset between X-ray centroid and BCG. 
Here we consider the location of our classes of regular/intermediate
and disturbed clusters, as determined from our $w-P_3$ classification
in the parameter planes of the $L_X-T$ and $M_X-T$ relations,
to study the power of the method by the significance with
which it reveals morphological segregation effects
in these relations. 

We model the scaling relation between the bolometric luminosity and
temperature for 35 clusters by $L/L_0$=$B (T/T_0)^A$ where
$L_0=10^{45}$~erg/s, and $T_0$=7~keV. We assume that the $L_X-T$ 
relation is non-evolving (see B\"ohringer et al. 2012), and do not use
a self-similar evolution. The luminosities used here are the values
from Zhang et al. (2008) without the cool-core correction, and
the temperatures used refer to the volume-averaged results. 
It is known that different fitting methods give very different 
results. We give the fitting results from BCES(Y$\big|$ X) 
(Akritas \& Bershady 1996) in Table 2. 
Okabe et al. (2010b) noted that
their results are very close to the results from the bisector, and we
give also these results by using the bisector fit in
Table 2.

\begin{table*}[]
  \begin{center}
    \begin{tabular}{r l l l l l l}
      \hline
      \hline
      \multicolumn{1}{l}{Relation } &
      \multicolumn{5}{c}{Fitting Method} \\
      \multicolumn{1}{c}{ } &
      \multicolumn{3}{c}{BCES (Y$|$X)} & \multicolumn{3}{c}{Bisector} \\
      \cline{2-7}
      \multicolumn{1}{c}{} & \multicolumn{1}{c}{A} & 
      \multicolumn{1}{c}{$B\,(10^{45}$ erg s$^{-1}$)} & 
      \multicolumn{1}{c}{$\sigma_{ln_{L}}$} &  \multicolumn{1}{c}{A} &
      \multicolumn{1}{c}{$B\,(10^{45}$ erg/s)} & 
      \multicolumn{1}{c}{$\sigma_{ln_{L}}$} \\
      \hline
      \\
      $L_X$--$T_X$ & 
      $2.58 \pm 0.59 $ & $1.32 \pm0.10$ & $0.50\pm0.07$ 
      & $3.44\pm 0.53$ & $1.29\pm0.10$ & $0.57\pm0.08$ \\ 
      (D)& 
      & $1.10\pm0.11$ &  & & $1.04\pm0.11$ & \\ 
      (R/I) & 
      & $1.54\pm0.15$ &  & & $1.55\pm 0.17$ & \\ 
      $L_XE(z)^{-1}$--$T_X$ & 
      $2.55 \pm 0.58 $ & $1.18\pm0.09$ & $0.50\pm0.07$ 
      & $3.41\pm 0.52$ & $1.16\pm0.09$ & $0.56\pm0.08$ \\ 
      $M_X$--$T_X$ & 
      $1.93 \pm 0.37 $ & $1.01\pm0.05$ & & $1.58\pm 0.47$ & $1.10\pm0.05$ & \\ 
      (D)& 
      & 1.19$\pm$0.08 &  & & 1.22$\pm$0.07& \\ 
      (R/I) & 
      & 1.01$\pm$0.05 &  & & 1.00$\pm$0.05 & \\ 
      \hline
    \end{tabular}
  \end{center}
  \caption{ {\footnotesize Results of the fitted scaling
      relations. $L_X$ is the corrected bolometric luminosity, $T$ is
      the volume-weighted temperatures, 
      as explained in Zhang et al. (2008). 
      For $L_x-T$ relation we give, for comparison,
      also the fitted results for self-similar evolution. 
      We also show the fitted normalisations with the fixed slope 
      of the $L_X-T$ relation for the two subsets of 35 clusters
      according to their morphological classification. 
      The results are given separately for the disturbed (D) 
      and for the regular/intermediate (R/I) clusters. The intrinsic
      scatter, $\sigma_{ln_{L}}$, measures the logarithmic scatter around
      the best fits. 
      No intrinsic scatter is given for $M_X-T$ because of 
      the large systematic errors in the mass determination.
    }
  }
\end{table*}

\begin{figure}
\begin{center}
\resizebox{\hsize}{!}{\includegraphics[height=6cm]{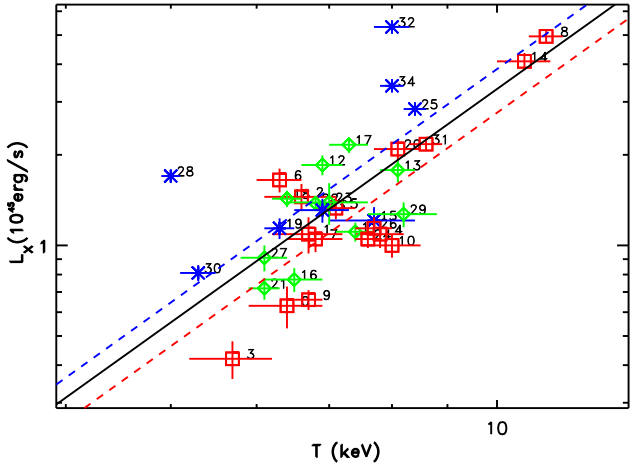}}
\end{center}
\caption{ 
  $L_X-T$ relation with the fitted results from the BCES(Y$\big|$X) method. 
  The black solid line is the best fit for all 35 clusters, the blue dashed 
  line is that for the regular and the intermediate clusters, and the red 
  dashed line for the disturbed clusters. The fitted normalisations 
  for the two sub-groups in blue and red move to the opposite directions 
  from the best fit found for all clusters. 
  The numbering of the clusters follows that of Fig. 1
}
\label{fig:lx-t-scaling}
\end{figure}

Our finding is shown in Fig.~\ref{fig:lx-t-scaling}. The best-fit for
all 35 clusters is shown as the black solid line. We divide the clusters into
two morphological groups, the regular and the intermediate together,
and the irregular clusters. The former consists of 19, and the latter
16 which divide the clusters into two groups of similar size. The
blue dashed line in Fig.~\ref{fig:lx-t-scaling} is the best fit for
regular/intermediate clusters, and the red dashed line is for
the irregular clusters. In both cases we fix the slope to the best
fit for 35 clusters, and only fit for the normalisations. We see a
trend that the more regular clusters have a higher, and the disturbed
clusters have a lower normalisation, both with 9--10 $\sigma$
significance. In the presence of cool cores one expects and has
previously observed such a trend as cool cores boost the luminosity 
(e.g. Fabian et al. 1994, Chen et al. 2007, Pratt et al. 2009),  
and cool cores are more frequent among the more regular clusters 
(Fig. 8 of B\"ohringer et al. 2010). This is why the trend is
found with a high significance since the X-ray luminosity fitted here
includes the cool core. 
We calculated the intrinsic scatter by the quadratic
difference between the scatter from the data and that expected from the
statistical measurement uncertainties with results listed in Table 2. 
This quantity should, however, be taken
with a caution since the measurement error includes systematics
which can not be corrected for in this type of calculation.

\begin{figure}
\begin{center}
\resizebox{\hsize}{!}{\includegraphics[height=6cm]{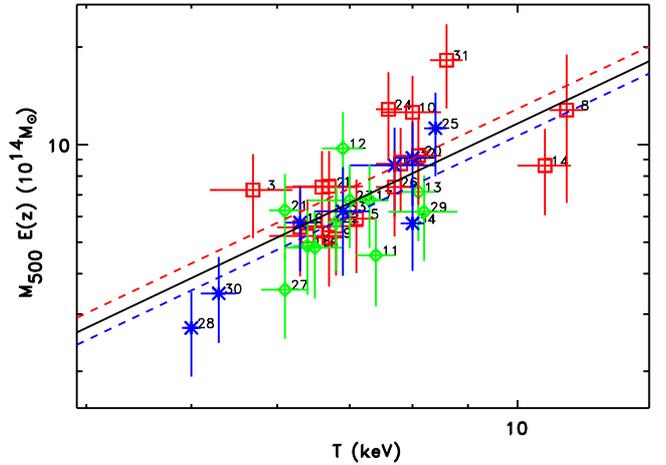}}
\end{center}
\caption{ 
  $M_X-T$ relation with the fitted results from the bisector method. 
  The fitted lines follow the same notation as in Fig.~\ref{fig:lx-t-scaling}. 
  The two morphological groups move to the opposite sides from the best-fit 
  of the 35 clusters. Disturbed clusters appear slightly more massive for 
  a given temperature.}
\label{fig:m-t-scaling}
\end{figure}

We show the scaling relation of $M_X-T$ with the best fitting model
function in Fig.~\ref{fig:m-t-scaling}. The cluster masses are taken
from Zhang et al. (2008), which were determined from the temperature
and density profiles derived from X-ray observations on the assumption
of hydrostatic equilibrium and spherical symmetry.
We model the relation by $M/M_0 E(z)$=$B
(T/T_0)^A$ where $M_0=6\times10^{14}M_{\sun}$ and $T_0$=7~keV, and
assume self-similar evolution. The results are given in Table 1 and
Fig.~\ref{fig:m-t-scaling}, where the notation is the same as in
Fig.~\ref{fig:lx-t-scaling}. When we fit the more regular and
disturbed subsamples separately with the slope fixed to that of
the total sample, we find a significantly higher normalisation
of the mass for a given temperature for the disturbed clusters.
The offset from the fit of the total sample is roughly 11\%
(-8.5\%) for the disturbed (more regular) clusters. 
The offset could possibly come from the fact that some of the 
energy in the ICM is in the form of kinetic energy of internal motions
which will subsequently be converted to heat when the cluster
relaxes. 
A substantial amount of turbulent energy is seen in post-merger clusters
in, for example, the observations of the Coma cluster (Schuecker et al. 2004)
and in simulations (Sunyaev et al. 2003). 
Thus in the relaxed phase the cluster would have a higher
temperature for a given mass. 
Alternatively the effect could come from an overestimation of
the cluster mass. The distortions of the cluster as seen 
in the surface brightness distribution most often lead 
to azimuthally averaged surface brightness and gas density profiles
with an extended core and steeper outer slope, both of which yield
an increase in the mass determination using spherical symmetry.
To further investigate if this interpretation is likely we also 
studied the $M_{\rm gas}-T$ relation for these clusters. 
For the $M_{\rm gas}-T$ relation we find offsets of similar sign as for
$M_X$, but the offsets are less than 1\%, much smaller than 
for $M_X$. As $M_{\rm gas}$ is a quantity that results from an integration 
over the gas density profile and not just from the slope
at $r_{500}$ as $M_X$, it is less prone to be affected by 
substructure. We therefore conclude that most likely the 
major part of the effect of disturbed clusters in the $M_X-T$ 
relation come from a mass overestimation due to the assumption
of hydrostatic equilibrium and spherical symmetry. 

\begin{figure}
\begin{center}
\resizebox{\hsize}{!}{\includegraphics[height=6cm]{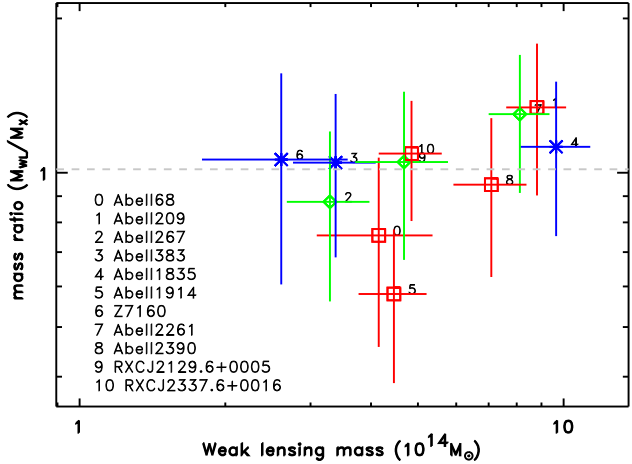}}
\end{center}
\caption{Ratio of the weak-lensing and X-ray determined masses as a
  function of the weak-lensing mass estimates. Morphologically
  disturbed clusters have the largest scatter compared to more regular
  clusters.}
\label{fig:mratio-lratio}
\end{figure}

A subsample of twelve clusters was previously analysed using the
weak-lensing and X-ray data by Okabe et al. (2010a,b) and Zhang et
al. (2010) defined by the overlapping sample between Subaru/Suprim-Cam
and \emph{XMM-Newton} observations in LoCuSS. In this section we study the
correlation between the X-ray morphologies presented in this
article, based on $w-P_3$, and the degree of agreement in the mass
estimation in the two different wavelengths. 
Our analysis contains currently eleven clusters which excludes 
Abell 115 for the reason explained above, and all mass estimates
are taken at $r_{500}$. A first look at the 
mass estimate comparison for clusters classified in 
the $w-P_3$ plane is provided in Fig.~\ref{fig:mratio-lratio}.  
We find that disturbed clusters present larger scatter in 
the mass comparison than regular/intermediate clusters. With 
a slight difference in classification for individual clusters
this result is in line with Zhang et al. where the X-ray 
morphology was determined by concentration and asymmetry 
parameters given in Okabe et al. This probably reflects the 
fact that mass determination is more difficult for those 
disturbed clusters.

\section[]{Summary and Conclusions}

For the \emph{XMM-Newton} observations of the 35 X-ray luminous galaxy
clusters we characterised the X-ray morphologies with the modified
power ratio and the centre shift method. The first moment that can
consistently describe the degree of departures from the symmetry is
the $P_3$, and together with the $w$-parameter we categorised the
clusters into three groups as regular, intermediate, and disturbed
after checking the consistency with other power ratio parameters. We
caution readers that this division is not unique and discrete, while
the distribution of parameters is rather continuous as was seen 
in section 4.1. Nevertheless this
classification also agrees with the visual impression of the X-ray
images and known properties of the clusters, and helps to
simplify the presentation of results.

Having a reliable description of the cluster morphology in hand, we investigated
the correlations of different cluster morphologies with a number of
cluster physical parameters. We took the X-ray luminosity ratio
between the core-included and the core-excised cases where a high
luminosity ratio is an indicator of a cool core in the cluster, and
compared to our representative structural parameters, $P_3$ and $w$.
Consistent with the earlier results in B\"ohringer et al. (2010) we
found that regular clusters have a larger luminosity ratio compared to
the morphologically disturbed clusters. This is also confirmed by
comparing the luminosity ratio to the $P_3$ and $w$ parameters, where
larger luminosity ratios are associated to low $w$ and $P_3$. This
result implies the known fact that cool cores are much more frequent
in regular clusters, and are rarely found in very disturbed systems. This
applies similarly to the central electron density, where we expect
that more X-ray emitting gas is concentrated at the centre of the
cool-core clusters.

Another interesting aspect studied was the comparison of the optical
and X-ray morphology. The luminosity gap between the BCG and the
second ranked cluster galaxy was compared with $P_3$, and
the general trend was found that the clusters with the most
dominant BCGs form a relatively homogeneous population of
regular/intermediate systems. In contrast, clusters with less
dominant BCGs present a more heterogeneous range of X-ray
morphologies including the most disturbed. These results are
consistent with the picture that the BCG dominance probes a combination
of cluster formation epoch and the time interval since the last
episode of major hierarchical infall into the cluster potential
well (Smith et al.\ 2010). Clusters that host dominant BCGs formed
relatively early, and have suffered negligible recent infall. This allows 
the bright galaxies to suffer dynamical friction and to merge to 
form a bright BCG, and the X-ray gas to settle down into a regular
morphology. On the other hand clusters with less dominant BCGs
either formed more recently, or formed relatively early but have
suffered more significant recent infall than their cousins that host
a dominant BCG.

We also compared our X-ray morphological analysis with the surface photometry 
of the relevant BCGs, specifically the ellipticity of these galaxies measured 
in projection on the sky. In general we find a good correlation between
the BCG ellipticity and the quadrupole moment of the X-ray surface brightness 
distribution, and a similar correlation with larger scatter between BCG 
ellipticity and the X-ray ellipticity. We thus find that the clusters that 
host relatively circular BCGs ($\epsilon_{\rm BCG}\textless0.2$) have 
exclusively small quadrupole moments and X-ray ellipticity. This is consistent 
with the suggestion by Marrone et al.'s (2012) that BCG ellipticity may be a
useful indicator of orientation of the underlying dark matter distribution 
of a cluster, with respect to the observer's line of sight where a small 
ellipticity is partly the result of the line-of-sight orientation of a prolate 
cluster and its BCG.

Finally, we examined the dependence of cluster locations in the $L_X-T$ and 
$M_X-T$ planes on their X-ray morphology, as determined by our $w-P_3$ 
classification, and also how well X-ray and lensing cluster mass estimates 
agree as a function of X-ray morphology. All three of these tests are based 
on previous published measurements of $L_X, T, M_X,$ and $M_{\rm WL}$, and 
can thus serve for consistency checks between our morphological analysis 
and complementary methods in the literature as well as for the study of 
new morphological diagnostics. We find that on average regular/intermediate 
clusters are separated from disturbed clusters in the $L_X-T$ 
plane, with the former 17\% more luminous than the latter at fixed
temperature, at $\sim9-10\sigma$ significance.
We stress that this
result is based on $L_X$ and $T$ measurements that do not excise
the core of each cluster. 
This result therefore demonstrates the
strong relationship between cooling timescale in cluster cores and
the location of clusters in the $w-P_3$ plane.
In our study of the $M_X-T$ relation we find that disturbed
clusters have a higher mass by $\sim 11$\% for a given temperature
than the overall average. We tentatively attribute most
of this effect to an overestimation of the mass for disturbed
clusters, since the corresponding $M_{\rm gas}-T$ relation does
not show such a strong segregation.
We also find that
disturbed clusters present larger scatter than regular/intermediate
clusters in the comparison on X-ray and lensing-based cluster mass
estimates, although the current sample is not large enough to draw
definitive conclusions.

In summary, this study together with our previous work
for the REXCESS cluster sample (B\"ohringer et al. 2010)
has shown that the morphological classification of galaxy
clusters by power ratios and centre shifts provides
an effective tool to analyse the influence 
of morphology on other properties of galaxy clusters.
We have demonstrated that these techniques
can be used to probe cluster morphology in a manner that correlates
well with other physical cluster properties, both at X-ray and
optical/near-infrared wavelengths. They therefore offer a promising
route to the morphological analysis of larger, more complete samples
that will form the basis of the cosmological interpretation of
upcoming large cluster samples from \emph{eROSITA}, and
\emph{EUCLID}.

\begin{acknowledgements} 
HB acknowledge support by the cluster of excellence ``Origin and
Structure of the Universe'' (www.universe-cluser.de). HB and GC
acknowledge support from the DfG Transregio Program TR 33. GC
acknowledges support from Deutsches Zentrum f\"ur Luft und
Raumfahrt(DLR). GPS acknowledges support from the Royal
Society, and thanks his colleagues in the LoCuSS collaboration for
helpful discussions.
\end{acknowledgements}

\appendix

\section[]{Morphological parameters for individual clusters}

In Table~\ref{tab:app} we provide the results for the morphological parameters,
power ratios, and centre shifts for the individual clusters in our
study sample. We also indicate the morphological type from our
classification based on both $w$ and $P_3$ as described in section 4.1

\begin{table*}
\begin{center}
\label{tab:app}
\centering
\begin{tabular}{l c c c c c c c c l}
\hline
\hline
\multicolumn{1}{l}{Cluster Name} & 
\multicolumn{1}{c}{$w$} & 
\multicolumn{1}{c}{$\sigma_w$} & 
\multicolumn{1}{c}{$P_2$} & 
\multicolumn{1}{c}{$\sigma_{P_2}$} & 
\multicolumn{1}{c}{$P_3$} & 
\multicolumn{1}{c}{$\sigma_{P_3}$} & 
\multicolumn{1}{c}{$P_4$} & 
\multicolumn{1}{c}{$\sigma_{P_4}$} & 
\multicolumn{1}{l}{Classification} \\ 
&  $\times10^{-2}$ & $\times10^{-2}$ & $\times10^{-5}$ & $\times10^{-5}$ & $\times10^{-5}$ & $\times10^{-5}$ & $\times10^{-5}$ & $\times10^{-5}$ & \\
\hline
Abell781&   2.230&   0.172&   0.788&   0.345&   0.440&   0.074&   0.517&   0.124&D\\
RXCJ0516.7-5430&   1.670&   0.140&   2.593&   0.440&   0.052&   0.126&  -0.044&   0.061&D\\
RXCJ0528.9-3927&   1.110&   0.060&   0.349&   0.111&   0.027&   0.011&   0.056&   0.010&I\\
RXCJ0945.4-0839&   0.928&   0.101&   0.438&   0.092&  -0.009&   0.009&   0.015&   0.009&D\\
RXCJ2234.5-3744&   0.854&   0.036&   0.664&   0.042&   0.008&   0.005&   0.006&   0.005&D\\
Abell209&   0.745&   0.038&   1.145&   0.091&   0.026&   0.009&   0.008&   0.004&D\\
Abell1763&   0.732&   0.052&   2.173&   0.117&  -0.001&   0.009&   0.008&   0.005&D\\
RXCJ0043.4-2037&   0.691&   0.059&   1.492&   0.300&  -0.008&   0.016&   0.123&   0.045&D\\
RXCJ0658.5-5556&   0.690&   0.018&   0.775&   0.056&   0.148&   0.014&   0.057&   0.006&D\\
RXCJ2218.6-3853&   0.657&   0.027&   2.964&   0.076&   0.014&   0.004&   0.010&   0.002&D\\
RXCJ2337.6+0016&   0.635&   0.047&   1.450&   0.169&   0.001&   0.016&  -0.001&   0.012&D\\
Abell2218&   0.603&   0.034&   0.641&   0.047&   0.019&   0.006&   0.006&   0.005&I\\
RXCJ0232.2-4420&   0.591&   0.040&   1.414&   0.065&   0.014&   0.005&   0.009&   0.002&I\\
RXCJ0645.4-5413&   0.555&   0.046&   0.841&   0.034&   0.013&   0.003&   0.020&   0.004&I\\
Abell2390&   0.553&   0.023&   1.068&   0.051&   0.012&   0.003&   0.015&   0.003&D(b)\\
RXCJ2308.3-0211&   0.477&   0.065&   0.318&   0.073&   0.007&   0.015&   0.003&   0.005&R\\
Abell267&   0.450&   0.060&   1.632&   0.130&   0.012&   0.014&   0.001&   0.005&I\\
Abell2667&   0.450&   0.045&   0.769&   0.039&   0.010&   0.004&   0.003&   0.001&I\\
RXCJ2129.6+0005&   0.420&   0.023&   0.958&   0.036&   0.010&   0.003&   0.009&   0.002&I\\
Abell963&   0.357&   0.028&   0.187&   0.026&   0.004&   0.002&   0.002&   0.001&R\\
Abell773&   0.341&   0.049&   0.623&   0.064&   0.028&   0.015&   0.011&   0.002&D\\
RXCJ0547.6-3152&   0.326&   0.020&   0.396&   0.028&   0.013&   0.004&   0.004&   0.002&I\\
Abell1413&   0.307&   0.018&   1.296&   0.032&   0.014&   0.003&   0.004&   0.001&I\\
Abell2261&   0.294&   0.067&   0.206&   0.066&   0.008&   0.010&   0.017&   0.006&I\\
Abell1758&   0.279&   0.079&   3.123&   0.129&   0.034&   0.016&   0.027&   0.004&D\\
Abell1689&   0.277&   0.009&   0.168&   0.006&   0.004&   0.001&   0.001&   0.000&R\\
Abell68&   0.232&   0.068&   2.194&   0.122&   0.029&   0.012&   0.008&   0.005&D\\
RXCJ0958.3-1103&   0.209&   0.036&   1.068&   0.074&   0.010&   0.006&   0.006&   0.003&I\\
Z7160&   0.193&   0.016&   0.282&   0.028&   0.003&   0.001&   0.014&   0.003&R\\
RXCJ0532.9-3701&   0.187&   0.040&   0.354&   0.063&   0.018&   0.014&   0.006&   0.007&I\\
Abell383&   0.187&   0.020&   0.016&   0.004&   0.006&   0.002&   0.014&   0.002&R\\
Abell1914&   0.166&   0.021&   0.745&   0.031&   0.041&   0.004&   0.007&   0.001&D\\
Abell1835&   0.136&   0.010&   0.123&   0.008&   0.003&   0.001&   0.005&   0.001&R\\
RXCJ0307.0-2840&   0.089&   0.031&   0.171&   0.051&   0.005&   0.010&   0.012&   0.007&R\\
Abell2204&   0.088&   0.013&   0.059&   0.004&   0.003&   0.001&   0.001&   0.000&R\\
\hline
\end{tabular}
\end{center}
\caption{
  Morphological classification of the 35 clusters in our sample according 
  to the revised centre shift and power ratios. D is short for disturbed, I 
  is for intermediate, and R is for regular clusters. Note that the negative 
  values in $P_3$ are the result of the bias correction. In this case the 
  significance of $P_3$ values is limited, and we mainly combine it with the 
  $w$-parameter to classify their morphological state after further checks 
  with the rest of the structural parameters.
}
\end{table*}

\end{document}